\documentclass[letterpaper]{article} % DO NOT CHANGE THIS
\usepackage{aaai2026}  % DO NOT CHANGE THIS
\usepackage{times}  % DO NOT CHANGE THIS
\usepackage{helvet}  % DO NOT CHANGE THIS
\usepackage{courier}  % DO NOT CHANGE THIS
\usepackage[hyphens]{url}  % DO NOT CHANGE THIS
\usepackage{graphicx} % DO NOT CHANGE THIS
\usepackage{natbib}  % DO NOT CHANGE THIS AND DO NOT ADD ANY OPTIONS TO IT
\usepackage{caption} % DO NOT CHANGE THIS AND DO NOT ADD ANY OPTIONS TO IT
\usepackage{algorithmic}
\usepackage{amsmath} 
\usepackage{amssymb}

\usepackage{multirow}
\usepackage[ruled,linesnumbered]{algorithm2e} 
\usepackage{booktabs}
\usepackage{newfloat}
\usepackage{listings}
\usepackage{tabularx}
\usepackage{makecell}
\DeclareCaptionStyle{ruled}{labelfont=normalfont,labelsep=colon,strut=off} % DO NOT CHANGE THIS
\title{Modeling Item-Level Dynamic Variability with Residual Diffusion for Bundle
Recommendation}
\author{
    Dong Zhang\textsuperscript{\rm 1,2}, Lin Li\textsuperscript{\rm 1}\thanks{Corresponding author}, Ming Li\textsuperscript{\rm 1,3}, Amran Bhuiyan\textsuperscript{\rm 3}, Meng Sun\textsuperscript{\rm 1}, Xiaohui Tao\textsuperscript{\rm 4}, Jimmy Xiangji Huang\textsuperscript{\rm 3}
}
\affiliations{
    \textsuperscript{\rm 1}School of Computer Science and Artificial Intelligence, Wuhan University of Technology, China\\
    \textsuperscript{\rm 2}Engineering Research Center of Hubei Province for Clothing Information, Wuhan Textile University, China\\
    \textsuperscript{\rm 3}Information Retrieval and Knowledge Management Research Lab, School of Information Technology, York University, Canada\\
    \textsuperscript{\rm 4}School of Mathematics, Physics and Computing,
University of Southern Queensland, Australia\\
    \{zhangdong23,cathylilin,liming7677,sunmeng\}@whut.edu.cn, amran.apece@gmail.com, Xiaohui.Tao@unisq.edu.au, jhuang@yorku.ca
}

\usepackage{bibentry}
\begin{document}

\maketitle

\begin{abstract}
Existing solutions for bundle recommendation (BR) have achieved remarkable effectiveness for predicting the user’s preference for prebuilt bundles. However, bundle-item (B-I) affiliation will vary dynamically in real scenarios. For example, a bundle themed as `casual outfit' may add `hat' or remove `watch' due to factors such as seasonal variations, changes in user preferences or inventory adjustments. Our empirical study demonstrates that the performance of mainstream BR models may fluctuate or decline under item-level variability. This paper makes the first attempt to address the above problem and proposes \underline{R}esidual \underline{Diff}usion for \underline{B}undle \underline{R}ecommendation (\textit{RDiffBR}) as a model-agnostic generative framework which can assist a BR model in adapting this scenario. During the initial training of the BR model, RDiffBR employs a residual diffusion model  to process the item-level bundle embeddings which are generated by the BR model to represent bundle theme via a forward-reverse process. In the inference stage, RDiffBR reverses item-level bundle embeddings obtained by the well-trained bundle model under B-I variability scenarios to generate the effective item-level bundle embeddings. In particular, the residual connection in our residual approximator significantly enhances BR models' ability to generate high-quality item-level bundle embeddings. Experiments on six BR models and four public datasets from different domains show that RDiffBR improves the performance of Recall and NDCG of backbone BR models by up to 23\%, while only increases training time about 4\%.
\end{abstract}

\begin{links}
    \link{Code}{https://github.com/WUT-IDEA/RDiffBR}
    \link{Extended version}{https://arxiv.org/abs/2507.03280}
\end{links}

\section{Introduction}

With the rapid development of online platforms, bundle recommendation (BR) has become an important tool for enhancing user experience and promoting sales. Specifically, users are provided with one-stop services by adding a group of related items to a bundle (such as booklists, playlists, fashion sets, and meal sets in takeout services) while improve sales efficiency and revenue. %At the same time, it allows users to simultaneously select multiple desired items in a single operation, thereby significantly reducing the time required for the selection process.

\begin{figure}[!t]
  \centering
  \includegraphics[width=\linewidth]{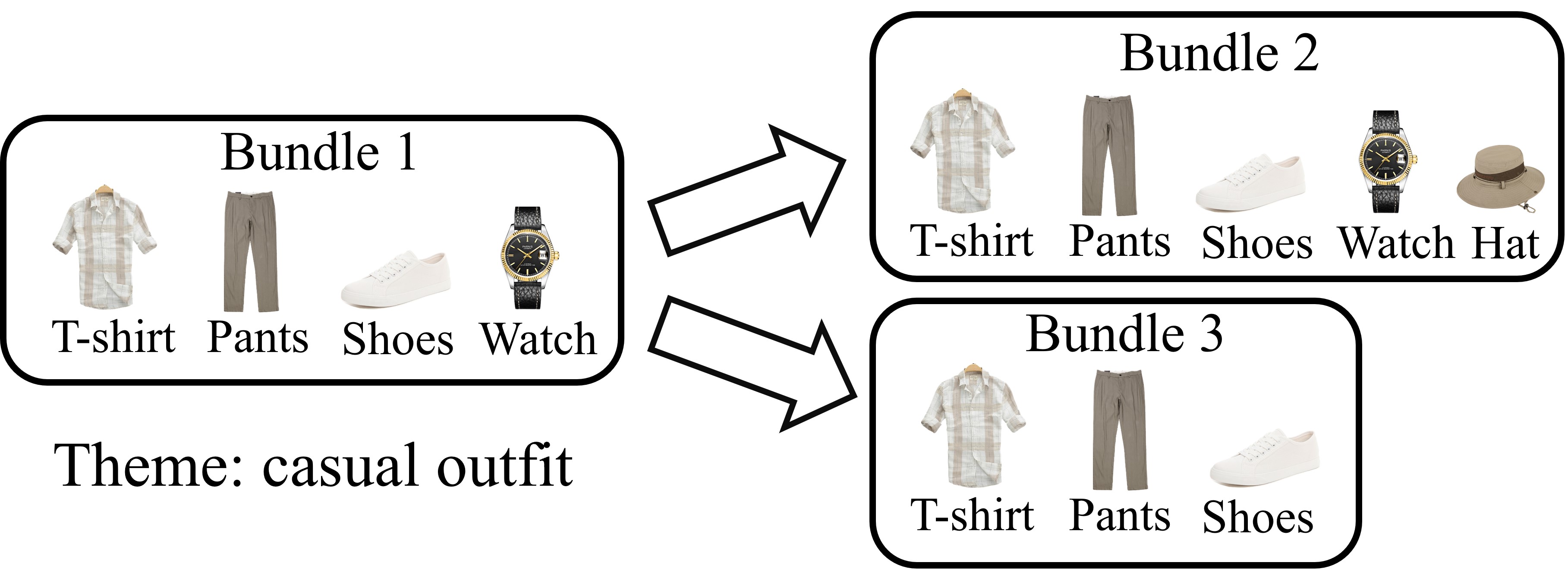}
  \caption{An example showing that bundle-item affiliation is changed while bundle theme remains relatively stable.}
  \label{figintro}
\end{figure}

Bundle recommendation usually learns from the user-bundle (U-B) interactions, user-item (U-I) interactions and bundle-item (B-I) affiliations to recommend a list of bundles to users. However, in real scenarios B-I affiliations may undergo dynamic changes due to factors such as seasonal variations, shifts in user preferences, or inventory adjustments, while the theme of the bundle remains. As illustrated in Figure~\ref{figintro}, Bundle $1$ initially includes `T-shirt', `pants', `shoes' and `watch' with the theme of `casual outfit'. 
In practice, platforms can add the `hat' to align with summer promotions or remove the `watch' based on user feedback. An intuitive idea is to regard it as a new bundle. However, such item-level changes may not definitely cause bundle theme shifting although bundle theme is jointly determined by the items within the bundle. As shown in Figure~\ref{figintro}, the theme of Bundle 2 and Bundle 3 remain `casual outfit' by item variation. It is advisable to give priority to reusing the identification information of the original bundle because this dynamic variability maintains the theme of bundle. 

In recent years, lots of works have focused on bundle recommendation and achieved good results. Including factorization models represented by LIRE~\cite{LIRE}, EFM~\cite{EFM}, attention-based models represented by DAM~\cite{DAM}, Attlist~\cite{AttList} and GNN-based models represented by CrossCBR~\cite{CrossCBR}, BundleGT~\cite{BundleGT} and DCBR~\cite{DCBR}. However, our experiments have found that the performance of existing mainstream bundle recommendation models will fluctuate or even decline under B-I dynamic variability scenarios we have mentioned above. For instance, we found that on Youshu dataset, Recall@20 of BGCN will decrease by up to 19\% as items in bundles increases or decreases (see Section 3.2 for details). This is because existing bundle recommendation models lack the ability to effectively model the item-level dynamic variability when generating bundle representations of item-level.

In this paper, we make the first attempt to address the issue of item-level dynamic variability in bundle recommendation and propose a novel \underline{R}esidual \underline{Diff}usion for \underline{B}undle \underline{R}ecommendation (\textit{RDiffBR}) model-agnostic generative framework to assist the model in adapting this scenario. During the initial training of the BR model, RDiffBR employs a residual diffusion model to process the item-level bundle embeddings which are generated by BR model to represent bundle theme. \textit{In the forward process}, RDiffBR injects controlled Gaussian noises into item-level bundle embeddings. \textit{In the reverse process}, RDiffBR iteratively removes noises by a designed residual approximator. RDiffBR jointly trains this residual approximator and BR model  via above forward-reverse process. \textit{In the inference stage}, to generate the effective item-level bundle embeddings, RDiffBR reverses item-level bundle embeddings obtained by the well-trained bundle model under B-I variability scenarios rather than using pure noise as the starting point and conduct bundle recommendation. Experiments on six bundle models and four public datasets from different domains show that RDiffBR significantly improves the performance of bundle models in dynamic B-I variability scenario. 

The main contributions of this work are summarized as follows. %\begin{itemize} 
First, we propose a generative residual diffusion framework for bundle recommendation to address item-level dynamic variability issue. Secondly, our residual approximator in diffusion can significantly enhance the item-level bundle embedding generation ability of BR models. Lastly, extensive experiments on four datasets from different domains show that RDiffBR enhances the performance of Recall and NDCG of the BR backbone models by up to 23\%, while only increasing the training time by about 4\%. 
\begin{figure*}[h]
  \centering
  \includegraphics[width=\linewidth]{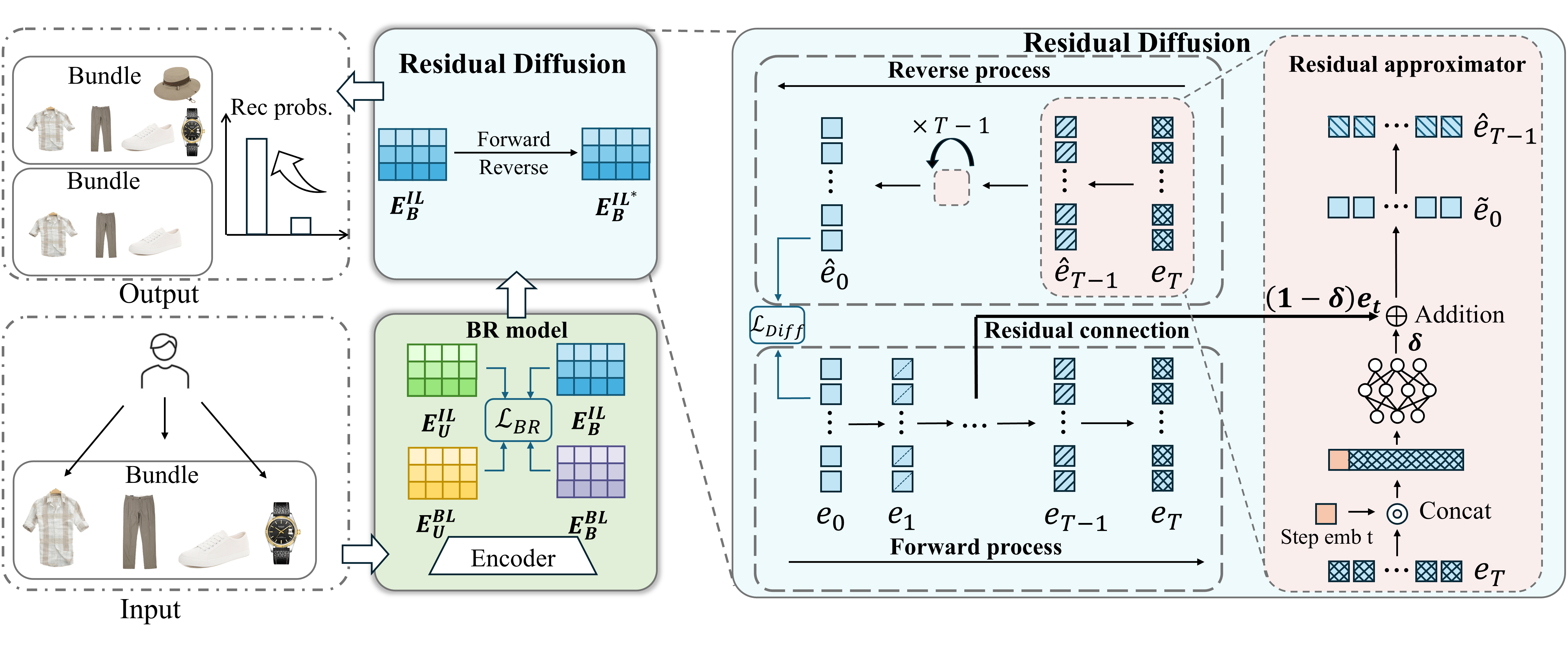}
  % \vspace{-10mm}
  \caption{Overview of the proposed RDiffBR framework.}
  \label{figFramework}
  % \vspace{-0.4cm}
\end{figure*}   
\section{Related Work}
\textbf{Bundle Recommendation.} Bundle recommendation aims to recommend a bundle of items related to a specific theme. Early work employed neural network methods for bundle recommendation~\cite{EFM,DAM,AttList,LIRE}. Recently, graph neural networks have become the mainstream technology for bundle recommendation. We can categorize these graph bundle recommendation models based on these three types of representation learning strategies~\cite{Survey_bundle}. $\bullet$ \textbf{Unified learning}. The unified representation learning strategy aims to learn representations of users and bundles within a unified user-bundle-item graph~\cite{POG,UHBR}. $\bullet$ \textbf{Separate learning}. The main idea of this strategy is to first perform representation learning and preference prediction upon two views individually, and then fuse two view-specific predictions to obtain the final prediction~\cite{BGCN,IHBR,BundleGT,HBGCN,CMRec}. $\bullet$ \textbf{Cooperative learning}. This approach learns user and bundle representations from bundle view and item view and uses techniques such as contrastive learning~\cite{MIDGN,CrossCBR,MultiCBR,HIDGN}, knowledge distillation~\cite{HyperMBR,DGMAE} or diffusion~\cite{DCBR} to achieve cross views mutual enhancement. However, most of bundle recommendation lack the ability to adapt to item-level dynamic variability. It is necessary to study our task for real-world application scenarios.

\textbf{Diffusion Models for Recommendation.} The emerging diffusion model in generative models, compared with GANs and VAEs, possess stronger stability and representation. Diffusion model plays three main roles in recommendation systems~\cite{survey_diffusion}: $\bullet$ \textbf{Diffusion for data augmentation}. This approach generating realistic samples to expand the original training data, such as sequence data~\cite{Diff4rec,Plug_diff} and side information~\cite{DiffKG,DiffMM}. $\bullet$ \textbf{Diffusion for representation enhancement}. Diffusion models converts the input data into embeddings to address some key challenges of recommendation. Lots of studies have been proposed such as denoising the implicit user feedback~\cite{DDRM}, sequential recommendation~\cite{DiffuRec}, out-of-distribution issue~\cite{CausalDiffRec} and multi-modal learning~\cite{MCDRec}. $\bullet$ \textbf{Diffusion as recommendation model}. This approach can be classified into three categories: collaborative recommendation~\cite{DRM,CFDiff}, context-aware recommendation~\cite{RecDiff,Diff-POI}, and other applications~\cite{DCDR}. However, a straightforward application of diffusion models on bundle recommendation is insufficient to address the issue of item-level dynamic variability. There are mainly two reasons: the loss targets of diffusion and bundle recommendation are inconsistent and the approximator needs to be designed based on item-level dynamic variability scenario.

\section{Methodology}

While it is intuitive to leverage side information—such as item attributes—to enrich the feature space, doing so typically demands non-trivial architectural changes to the original model. In contrast, our RDiffBR operates without any external information and leaves the base architecture untouched. The technical details are presented below.
\subsection{Problem Definition}
$ U (|U|=M)$ is denoted as the user set, $B (|B|=N) $as the bundle set, and $I (|I|=O)$ as the item set. The user-bundle interactions, user-item interactions, and bundle-item affiliations are represented as $X_{M\times N}=\{(u,b)\mid u\in U,b\in B\}$, $Y_{M\times O}=\{(u,i)\mid u\in U,i\in I\}$ and $Z_{N\times O}=\{(b,i)\mid b\in B,i\in I\}$ with a binary value at each entry respectively. The value 1 indicates an observed interaction of user-item or user-bundle, or affiliation of bundle-item. The item-level dynamic variability problem can be formulated as follows:

\textbf{Input}: The user-bundle interactions $X_{M\times N}$, user-item interactions $Y_{M\times O}$, and bundle-item affiliations $Z_{N\times O}$.

\textbf{Output}: A well-trained bundle recommendation estimates the probability $\hat{y}_{ub}\in(0,1)$ of a user $u$ interacting with unseen bundles $b$ based on $Z^*=Z\cup\Delta^+\setminus \Delta ^-$ rather than $Z$. Here, $\Delta^+=\{(b,i)\mid(b,i)\notin Z \} $ and $\Delta^-=\{(b,i)\mid(b,i)\in Z \}$ represent dynamic variability bundle-item affiliation. When an item is replaced by others, this can be treated as the joint operation of $\Delta^+$ and  $\Delta^-$.
\subsection{RDiffBR Framework}
Considering the successful experience of diffusion as a generative model and the sparsity of input data, our RDiffBR framework introduces a residual diffusion model in the latent space of interactions. As shown in Figure~\ref{figFramework}, after BR model utilizes an encoder to get bundle-level user/bundle embeddings $E^{BL}_u ,E^{BL}_b$ and item-level embeddings user/bundle embeddings $E^{IL}_u,E^{IL}_b$ from $\{X, Y, Z\}$. Since $E^{IL}_b$ represents the bundle theme generated by the BR model, we iteratively process $E^{IL}_b$ through a forward-backward process, jointly optimizing the BR model and the residual approximator module that we have designed based on the task. In the inference stage, RDiffBR applies this optimized forward-reverse process to $E^{IL}_b$ from dynamic variability B-I affiliations $Z^*$ to obtain new $E^{IL}_b$, % for bundle recommendation, 
enhancing BR model's ability for this scenario.
\subsubsection{\textbf{Residual Diffusion}}
\
\newline
$\bullet$ \textbf{Forward process.}
Given item-level bundle embeddings $E^{IL}_b$ from a BR model's encoding we can obtain a bundle's item-level bundle embedding $e^{IL}_b$ and set $e_0 = e^{IL}_b$. As shown in the bottom of residual diffusion module in Figure~\ref{2}, Gaussian noises are continuously incorporated into $e_0$ with adjustable scales and steps:
\begin{equation}\label{1}q(e_t|e_{t-1})=\mathcal{N}(e_t;\sqrt{1-\beta_t}e_{t-1},\beta_t\boldsymbol{I}).\end{equation}
Based on the reparameterization~\cite{Diffusion} and the additivity of  independent Gaussian noises, we can directly obtain $e_t$ from $e_0$.
\begin{equation}\label{2}q(e_t|e_0)=\mathcal{N}(e_t,\sqrt{\bar{\alpha}_t}e_0,(1-\bar{\alpha_t})\boldsymbol{I}),\end{equation}
where $\bar{\alpha_{t}}=\prod_{t=1}^{t}\alpha_{t},\alpha_{t}=1-\beta_{t},\beta_{t}\in(0,1)$ controls the Gaussian noise scales added at each step $t$. To regulate the added noises in $e_{1:T}$, we follow ~\cite{DRM} using a linear variance noise schedule in the forward process:
\begin{equation}\label{3}1-\bar{\alpha}_t=s\cdot\left[\alpha_{\min}+\frac{t-1}{T-1}(\alpha_{\max}-\alpha_{\min})\right],\end{equation}
where T is the total forward step and s $\in$ (0, 1) controls the noise scale. The condition $\alpha_{min}$ $<$ $\alpha_{max}$ $\in$ (0, 1) indicates the upper and lower limits of the added noise.
\\$\bullet$ \textbf{Reverse process.}
After obtaining the noisy embedding $e_T$ by forward process, we iteratively denoise this embedding in reverse processing as shown in the top of residual diffusion module in Figure~\ref{2}. In each reverse step, a residual approximator module to denoise them.
\begin{equation}\label{4}p_\theta(\hat{e}_{t-1}|\hat{e}_t)=\mathcal{N}(\hat{e}_{t-1};\mu_\theta(\hat{e}_t,t),\Sigma_\theta(\hat{e}_t,t)),\end{equation}
where $\hat{e}_t$ is denoised embedding in the reverse step t, $\theta$ is the learnable parameter of residual approximator. This module is executed iteratively in the reverse process until reconstructing final embedding $\hat{e}_0$
\\$\bullet$ \textbf{Residual approximator.}
We design a residual approximator module based on our task to denoise the noisy embedding in each reverse step as shown in the right of residual diffusion module. The diffusion training aims to approximate the distribution $ p_\theta(\hat{e}_{t-1}|\hat{e}_t)$ with the tractable posterior distribution $q(e_{t-1}|e_t, e_0 )$ in the reverse process. Through Bayes’ theorem, we can derive:
\begin{equation}\label{5}\begin{aligned}&q(e_{t-1}|e_t,e_0)\propto\mathcal{N}(e_{t-1};\tilde{\mu}(e_t,e_0,t),\sigma^2(t)\boldsymbol{I}),\mathrm{~where}\\&\begin{cases}\tilde{\mu}(e_t,e_0,t)=\frac{\sqrt{\alpha_t}(1-\bar{\alpha}_{t-1})}{1-\bar{\alpha}_t}e_t+\frac{\sqrt{\bar{\alpha}_{t-1}}(1-\alpha_t)}{1-\bar{\alpha}_t}e_0\\\sigma^2(t)=\frac{(1-\alpha_t)(1-\bar{\alpha}_{t-1})}{1-\bar{\alpha}_t},&\end{cases}\end{aligned}\end{equation}
$\tilde{\mu}(e_t, e_0,t)$ and $\sigma^2(t)I$ are the mean and covariance of $q(e_{t-1}|e_t, e_0 )$.  Following ~\cite{DDRM} we can similarly factorize $p_\theta (e_{t-1}|e_t )$:
\begin{equation}\label{6}\begin{aligned}&p_\theta(\hat{e}_{t-1}|\hat{e}_t)=\mathcal{N}(\hat{e}_{t-1};\mu_\theta(\hat{e}_t,t),\tilde{\beta}_t\boldsymbol{I})),\quad\mathrm{~where}\\&\mu_{\theta}(\hat{e}_{t},t)=\frac{\sqrt{\alpha_{t}}(1-\bar{\alpha}_{t-1})}{1-\bar{\alpha}_{t}}\hat{e}_{t}+\frac{\sqrt{\bar{\alpha}_{t-1}}(1-\alpha_{t})}{1-\bar{\alpha}_{t}}\tilde{e}_{0},\end{aligned}\end{equation}
$\tilde{e}_{0}$ is the predicted $e_0$ since the distribution of $e_0$ is unknown in the reverse process. We employ the multi-layer perceptron (MLP) to reconstruct $e_0$ in the  residual approximator. Specifically, for  embedding $e_t$ in the reverse step t:
\begin{equation}\label{7}\tilde{e}_0=\delta  f_\theta(\hat{e}_t,t)+(1-\delta )e_{t'},\end{equation}
where $\tilde{e}_0$ is the predicted value of $e_0$ by  residual approximator with parameter $\theta$. $\delta$ is residual connection weight. $e_{t'}$ is noisy embedding in the forward step $t'$. Step information t is encoded by sinusoidal positional encoding~\cite{Diffusion}, and these two inputs are concated as input.
\subsubsection{\textbf{Loss Function.}}
To optimize the forward-reverse process, it is necessary to minimize the variational lower bound of the predicted embedding. According to the KL divergence based on the multivariate Gaussian distribution, the reconstruction loss of within one training iteration can be expressed as:
\begin{equation}\label{8}\begin{aligned}\mathcal{L}_{r}&=\mathbb{E}_{q(\mathbf{e}_t|\mathbf{e}_0)}\left[D_{\mathrm{KL}}\left(q\left(\mathbf{e}_{t-1}|\mathbf{e}_t,\mathbf{e}_0\right)\|p_\theta\left(\mathbf{e}_{t-1}|\mathbf{e}_t\right)\right)\right]\\&=\mathbb{E}_{q(\mathbf{e}_t|\mathbf{e}_0)}\left[\frac{1}{2\sigma^2(t)}\left[\|\mu_\theta(\mathbf{e}_t,t)-\mu(\mathbf{e}_t,\mathbf{e}_0,t)\|_2^2\right]\right].\end{aligned}\end{equation}
Furthermore, by substituting Eq.\eqref{5} and Eq.\eqref{6} into Eq.\eqref{8} and simplifying~\cite{DRM} we can drive:
\begin{equation}\label{9}\mathcal{L}_\mathrm{diff}=\mathbb{E}_{t\sim\mathcal{U}(1,T)}\mathbb{E}_{q}\left[||e_{0}-\hat{e}_{\theta}(e_{t},e_{t'},t)||_{2}^{2}\right],\end{equation}
where $\mathcal{U}$ means uniformly sampling t from $\{1, 2, ...,T \}$. The final loss function of RDiffBR consists of: a bundle model loss and a reconstruction loss for reverse process. We have designed a loss balancing factor $\lambda$ to adjust the weight of reconstruction loss and jointly train the bundle model and  residual approximator together:
\begin{equation}\label{10}\mathcal{L}=\mathcal{L}_{BR\mathrm{~model}}+\lambda\mathcal{L}_{\mathrm{diff}}.\end{equation}
The training step of residual approximator is illustrated in Algorithm 1. Specifically, Lines 2-5 represent the forward process, Line 6 describes the reverse process using residual approximator, and Lines 7-8 address optimization.
\begin{algorithm}[!ht]

\caption{Residual approximator training process}
\KwIn{interaction data $X$, item-level bundle embedding $e_0$, diffusion step $T$, randomly initialized residual approximator MLP $f_{\theta}$.}
        \Repeat{converged}{
        Sample a batch of U-B interactions $\bar{X}\subset X$.\\
        
        \For{$all\ (u,b)\in\bar{X}$ }{
        Sample $t\sim\mathcal{U}(1,T)$;\\
        
        Compute $e_t$ given $e_0$ and $t$ via $q(e_t|e_0)$ in Eq.\eqref{2};\\
        
        Reconstruct $\tilde{e}_0$ through $f_{\theta}$ in Eq.\eqref{7};
        
        Calculate $\mathcal{L}$ by Eq.\eqref{10};  \\
        
        Take gradient descent step on $\nabla_{\theta}(\mathcal{L})$ to optimize $\theta$;  
        }  
}
\KwOut{optimized $\theta.$}
\end{algorithm}

\subsubsection{\textbf{Inference.}} In the inference stage, we take the item-level bundle embedding $e^*_0$ in the B-I dynamic variability scenario as input. Subsequently, RDiffBR continuously introduces noise into $e^*_0$ in the sequence $e^*_0 \to e^*_1\to ...\to e^*_{T'}$ for $T'$ steps in the forward process and then we set $\hat{e}_{{T'}}=e^*_{T'}$ to execute reverse process $ \hat{e_{T'}}\to\hat{e}_{{T'}-1} \to ... \to \hat{e}_{0}$ to generate new item-level bundle embedding $\hat{e}_0$. The reverse ignores the variance and uses $\hat{e}_{t-1} = \mu_\theta(\hat{e}_t,e^*_0,t)$ for directly inference following~\cite{MultiVAE}. We replace the original $e^*_0$ with $\hat{e}_{0}$, and then conduct bundle recommendation following the original way of the bundle model. Finally the ranking of bundles is enhanced. The inference procedure of residual approximator is illustrated in Algorithm 2. Specifically, Line 1 represent the forward process, Lines 2-5 represent the reverse process using a residual approximator to obtain ideal item-level bundle embedding.

\begin{algorithm}[!ht]
\caption{Inference process}

\KwIn{item-level bundle embedding $e_0^*$ from B-I dynamic variability scenario, sampling step $T^{'}$, well-trained residual approximator MLP $f_{\theta}$.}

    Compute $e_{T^{'}}^*$ given $e_0^*$ and $T^{'}$ via $q(e_{T^{'}}^*|e_0^*)$ in Eq.\eqref{2}.\\
    
    \For{$t = T^{'},...,1$}{
    Reconstruct $\tilde{e}_{0}$ through $f_{\theta}$ in Eq.\eqref{7};
    
    Compute $\hat{e}_{t-1}$ from $\hat{e}_{t}$ and $\tilde{e}_{0}$ in Eq.\eqref{6};
    
    }
           
\KwOut{item-level bundle embedding $e_0$.}  
\end{algorithm}

\section{Experiments}

Our experiments aim to answer the following questions:
\\$\bullet$\textbf{ RQ1:} How does RDiffBR perform with backbone bundle models under various degrees of B-I variability?
\\$\bullet$\textbf{ RQ2:} What is the contribution of residual connection in RDiffBR to its overall performance?
%\\$\bullet$\textbf{RQ3:} What is the effect of hyper-parameters on the result of RDiffBR?
%\\$\bullet$\textbf{RQ4:}
\\$\bullet$\textbf{ RQ3:} How does the efficiency of RDiffBR?
\subsection{Experimental Settings}
$\bullet$ \textbf{Datasets.}
Four publicly bundle datasets from different domains are summarized in Table~\ref{tab:Dataset Statistics}. Youshu~\cite{DAM} for book list recommendation; NetEase~\cite{EFM} for music playlist recommendation; iFashion~\cite{POG} for fashion outfit recommendation; In addition, we introduce another meal recommendation dataset $\mathrm{MealRec+}$~\cite{mealrec+}, where the meal consisted of individual course items is treated as bundle.
% \begin{table}[h]
% \Huge
%   \centering
  
%     % \vspace{-2mm}
%     \resizebox{\linewidth}{!}{
%     \begin{tabular}{ccccccccc}
 
%     \toprule
%     Dataset  & \#U   & \#I   & \#B   &  \#U-I  & \#U-B  & \#B-I  & \#Avg.I/B & \#Dens.B-I\\
%     \midrule
%     Youshu  & 8,039 & 32,770 & 4,771 & 138,515 & 51,377 & 176,667 & 37.03 &0.11\% \\
%     NetEase & 18,528 & 123,628 & 22,864 & 1,128,065 & 302,303 & 1,778,838 & 77.8 &0.06\%\\
%     iFashion  & 53,897 & 42,563 & 27,694 & 2,290,645 & 1,679,708 & 106,916 & 3.86 &0.01\% \\
%     MealRec$\mathrm{+}$& 1,575 & 7,280 & 3,817 & 151,148 & 46,767 & 11,451 & 3 &0.77\% \\
%     \bottomrule
%     \end{tabular}%
%     }
%    \caption{Dataset Statistics.} 
%   \label{tab:Dataset Statistics}%
%   % \vspace{-2mm}
% \end{table}%

\begin{table}[t]
\footnotesize
\centering
\label{tab:statistics}
\begin{tabularx}{0.47\textwidth}{lcccc}
\toprule 
Dataset          &Youshu  &NetEase  &iFashion   & MealRec$^+_H$ \\ 

\hline
\# U               & 8,039    & 18,528 & 53,897 &  1,575   \\
\# I               & 32,770  & 123,628 &  42,563 & 7,280  \\
\# B               & 4,771    & 22,864 & 27,694 &  3,817   \\
\# U-I              & 138,515    & 1,128,065 & 2,290,645 &151,148 \\
\# U-B              & 51,377    & 302,303 & 1,679,708 &  46,767   \\
\# B-I               & 176,667    & 1,778,838 & 106,916 &  11,451   \\
\# Avg. I/B                & 37.03    & 77.8  & 3.86 &  3   \\
Dens.B-I               & 0.11\%    & 0.06\% & 0.01\% &  0.77\%    \\ [-2pt]
\bottomrule
\end{tabularx}
\caption{Statistics of datasets.}
\label{tab:Dataset Statistics}
\end{table}
\begin{table*}[!t]
  \centering
\renewcommand{\arraystretch}{0.6} 

\setlength{\tabcolsep}{1.3pt}
    \small
     \begin{tabular}{l|cc|cc|cc|cc|cc|cc|cc|cc|cc|cc}
   
    \toprule
    Dataset & \multicolumn{10}{c|}{\textbf{Youshu}}                                         & \multicolumn{10}{c}{\textbf{NetEase}} \\[-2pt]
     \midrule
    $\rho$  & \multicolumn{2}{c|}{-4} & \multicolumn{2}{c|}{-2} & \multicolumn{2}{c|}{0} & \multicolumn{2}{c|}{2} & \multicolumn{2}{c|}{4} & \multicolumn{2}{c|}{-4} & \multicolumn{2}{c|}{-2} & \multicolumn{2}{c|}{0} & \multicolumn{2}{c|}{2} & \multicolumn{2}{c}{4} \\[-2pt]
     \midrule
    Metrics & R20   & N20   & R20   & N20   & R20   & N20   & R20   & N20   & R20   & N20   & R20   & N20   & R20   & N20   & R20   & N20   & R20   & N20   & R20   & N20 \\[-2pt]
     \midrule

    BGCN  & .1897  & .1047  & .2254  & .1278  & .2367  & .1380  & .2214  & .1247  & .2138  & .1204  & .0581  & .0309  & .0634  &.0333  & .0643  & .0339  & .0646  & .0341  & .0622  & .0332  \\ %[-10pt]
    +RDiffBR & .2178  & .1278  & .2391  & .1415  & .2453  & .1466  & .2470  & .1423  & .2407  & .1396  & .0614  & .0321  & .0657  & .0340  & .0662  & .0343  & .0668  & .0345  & .0649  & .0340  \\
    \textbf{\%improv.} & \textbf{14.83} & \textbf{22.15} & \textbf{6.06} & \textbf{10.74} & \textbf{3.61} & \textbf{6.25} & \textbf{11.57} & \textbf{14.11} & \textbf{12.60} & \textbf{16.03} & \textbf{5.56} & \textbf{3.98} & \textbf{3.63} & \textbf{1.99} & \textbf{2.89} & \textbf{1.14} & \textbf{3.47} & \textbf{1.07} & \textbf{4.42} & \textbf{2.54} \\[-1pt]

\midrule
    MIDGN & .2388  & .1373  & .2557  & .1472  & .2598  & .1496  & .2578  & .1495  & .2566  & .1489  & .0540  & .0278  & .0625  & .0330  & .0636  & .0337  & .0639  & .0337  & .0640  & .0337  \\
    +RDiffBR & .2498  & .1441  & .2621  & .1498  & .2645  & .1515  & .2652  & .1517  & .2643  & .1516  & .0568  & .0293  & .0649  & .0340  & .0661  & .0347  & .0667  & .0348  & .0672  & .0350  \\
    \textbf{\%improv.} & \textbf{4.62} & \textbf{4.98} & \textbf{2.53} & \textbf{1.72} & \textbf{1.79} & \textbf{1.32} & \textbf{2.87} & \textbf{1.49} & \textbf{3.00} & \textbf{1.82} & \textbf{5.21} & \textbf{5.27} & \textbf{3.89} & \textbf{2.91} & \textbf{3.96} & \textbf{2.92} & \textbf{4.45} & \textbf{3.25} & \textbf{4.95} & \textbf{4.14} \\[-1pt]
    \midrule
    HyperMBR & .1464  & .0953  & .2126  & .1290  & .2649  & .1581  & .2167  & .1285  & .1897  & .1162  & .0588  & .0313  & .0675  & .0354  & .0717  & .0377  & .0689  & .0363  & .0661  & .0345  \\
    +RDiffBR & .1789  & .1109  & .2229  & .1327  & .2669  & .1594  & .2280  & .1347  & .2128  & .1275  & .0627  & .0332  & .0694  & .0369  & .0739  & .0386  & .0704  & .0369  & .0667  & .0350  \\
    \textbf{\%improv.} & \textbf{22.20} & \textbf{16.42} & \textbf{4.85} & \textbf{2.87} & \textbf{0.75} & \textbf{0.83} & \textbf{5.20} & \textbf{4.87} & \textbf{12.14} & \textbf{9.75} & \textbf{6.67} & \textbf{5.97} & \textbf{2.88} & \textbf{4.41} & \textbf{3.07} & \textbf{2.51} & \textbf{2.25} & \textbf{1.79} & \textbf{0.85} & \textbf{1.50} \\[-1pt]
    \midrule
    CrossCBR & .2677  & .1584  & .2763  & .1626  & .2776  & .1640  & .2788  & .1649  & .2800  & .1656  & .0715  & .0386  & .0776  & .0419  & .0794  & .0431  & .0802  & .0436  & .0798  & .0434  \\
    +RDiffBR & .2744  & .1640  & .2814  & .1659  & .2827  & .1663  & .2816  & .1664  & .2816  & .1662  & .0772  & .0419  & .0818  & .0436  & .0826  & .0443  & .0829  & .0444  & .0826  & .0445  \\
    \textbf{\%improv.} & \textbf{2.52} & \textbf{3.57} & \textbf{1.84} & \textbf{2.02} & \textbf{1.85} & \textbf{1.42} & \textbf{1.01} & \textbf{0.92} & \textbf{0.56} & \textbf{0.40} & \textbf{7.99} & \textbf{8.50} & \textbf{5.36} & \textbf{4.04} & \textbf{4.09} & \textbf{2.67} & \textbf{3.43} & \textbf{1.70} & \textbf{3.50} & \textbf{2.51} \\[-1pt]
    \midrule
    MultiCBR & .2770  & .1649  & .2769  & .1649  & .2791  & .1652  & .2767  & .1641  & .2770  & .1648  & .0843  & .0463  & .0867  & .0472  & .0876  & .0475  & .0880  & .0479  & .0879  & .0479  \\
    +RDiffBR & .2831  & .1674  & .2840  & .1677  & .2824  & .1673  & .2825  & .1669  & .2842  & .1675  & .0859  & .0469  & .0879  & .0478  & .0881  & .0480  & .0883  & .0482  & .0893  & .0487  \\
    \textbf{\%improv.} & \textbf{2.20} & \textbf{1.55} & \textbf{2.56} & \textbf{1.71} & \textbf{1.19} & \textbf{1.25} & \textbf{2.10} & \textbf{1.71} & \textbf{2.60} & \textbf{1.61} & \textbf{1.91} & \textbf{1.32} & \textbf{1.37} & \textbf{1.27} & \textbf{0.57} & \textbf{1.18} & \textbf{0.40} & \textbf{0.75} & \textbf{1.60} & \textbf{1.52} \\[-1pt]
    \midrule
    BundleGT & .2687  & .1590  & .2872  & .1683  & .2877  & .1694  & .2873  & .1697  & .2837  & .1679  & .0729  & .0367  & .0869  & .0465  & .0874  & .0468  & .0874  & .0468  & .0871  & .0465  \\
    +RDiffBR & .2829  & .1653  & .2927  & .1708  & .2922  & .1707  & .2939  & .1712  & .2903  & .1702  & .0759  & .0384  & .0900  & .0479  & .0896  & .0479  & .0897  & .0477  & .0895  & .0476  \\
    \textbf{\%improv.} & \textbf{5.27} & \textbf{3.96} & \textbf{1.90} & \textbf{1.49} & \textbf{1.57} & \textbf{0.79} & \textbf{2.30} & \textbf{0.93} & \textbf{2.31} & \textbf{1.40} & \textbf{4.06} & \textbf{4.75} & \textbf{3.49} & \textbf{3.16} & \textbf{2.52} & \textbf{2.42} & \textbf{2.67} & \textbf{1.86} & \textbf{2.76} & \textbf{2.37} \\[1pt]

      \toprule

    Dataset & \multicolumn{10}{c|}{\textbf{iFashion}}                                                   & \multicolumn{10}{c}{\textbf{MealRec$^+_H$}} \\[-1pt]
    \midrule
    $\rho$  & \multicolumn{2}{c|}{-4} & \multicolumn{2}{c|}{-2} & \multicolumn{2}{c|}{0} & \multicolumn{2}{c|}{2} & \multicolumn{2}{c|}{4} & \multicolumn{2}{c|}{-4} & \multicolumn{2}{c|}{-2} & \multicolumn{2}{c|}{0} & \multicolumn{2}{c|}{2} & \multicolumn{2}{c}{4} \\[-2pt]
    \midrule
    Metrics & R20   & N20   & R20   & N20   & R20   & N20   & R20   & N20   & R20   & N20   & R20   & N20   & R20   & N20   & R20   & N20   & R20   & N20   & R20   & N20 \\[-2pt]
    \midrule
    BGCN  & .0473  & .0331  & .0560  & .0395  & .0602  & .0426  & .0607  & .0428  & .0604  & .0427  & .1068  & .1888  & .1213  & .2094  & .1325  & .2247  & .1326  & .2250  & .1375  & .2323  \\
    +RDiffBR & .0488  & .0341  & .0568  & .0398  & .0606  & .0428  & .0608  & .0429  & .0607  & .0429  & .1180  & .2078  & .1293  & .2180  & .1408  & .2401  & .1387  & .2371  & .1388  & .2384  \\
    \textbf{\%improv.} & \textbf{3.09} & \textbf{2.91} & \textbf{1.55} & \textbf{0.58} & \textbf{0.61} & \textbf{0.45} & \textbf{0.16} & \textbf{0.23} & \textbf{0.57} & \textbf{0.47} & \textbf{10.07} & \textbf{10.55} & \textbf{4.08} & \textbf{6.53} & \textbf{6.82} & \textbf{6.26} & \textbf{5.38} & \textbf{4.57} & \textbf{2.63} & \textbf{0.92} \\[-1pt]
    \midrule
    MIDGN & .0372  & .0257  & .0549  & .0380  & .0672  & .0482  & .0428  & .0279  & .0441  & .0294  & .0820  & .1564  & .1017  & .1816  & .1121  & .1908  & .1129  & .1919  & .1115  & .1880  \\
    +RDiffBR & .0426  & .0290  & .0628  & .0437  & .0763  & .0553  & .0534  & .0348  & .0545  & .0363  & .0826  & .1604  & .1075  & .1930  & .1179  & .2050  & .1145  & .2012  & .1146  & .2027  \\
    \textbf{\%improv.} & \textbf{14.55} & \textbf{12.83} & \textbf{14.36} & \textbf{14.87} & \textbf{13.53} & \textbf{14.65} & \textbf{24.66} & \textbf{24.83} & \textbf{23.48} & \textbf{23.77} & \textbf{2.60} & \textbf{0.82} & \textbf{6.29} & \textbf{5.69} & \textbf{7.44} & \textbf{5.13} & \textbf{4.87} & \textbf{1.48} & \textbf{7.82} & \textbf{2.79} \\[-1pt]
    \midrule
    HyperMBR & .0448  & .0336  & .0687  & .0494  & .0815  & .0580  & .0773  & .0545  & .0744  & .0521  & .0859  & .1498  & .0939  & .1595  & .1069  & .1778  & .1057  & .1781  & .1106  & .1887  \\
    +RDiffBR & .0462  & .0349  & .0692  & .0504  & .0830  & .0595  & .0797  & .0567  & .0773  & .0550  & .0874  & .1585  & .0995  & .1760  & .1122  & .1944  & .1205  & .1988  & .1247  & .2144  \\
    \textbf{\%improv.} & \textbf{3.21} & \textbf{4.09} & \textbf{0.81} & \textbf{2.01} & \textbf{1.75} & \textbf{2.69} & \textbf{3.01} & \textbf{4.01} & \textbf{3.92} & \textbf{5.42} & \textbf{5.82} & \textbf{1.70} & \textbf{10.33} & \textbf{5.90} & \textbf{9.34} & \textbf{4.94} & \textbf{11.61} & \textbf{14.00} & \textbf{13.62} & \textbf{12.82} \\[-1pt]
    \midrule
    CrossCBR & .0624  & .0491  & .0877  & .0683  & .0999  & .0772  & .1046  & .0799  & .1081  & .0831  & .0612  & .1048  & .0819  & .1417  & .0938  & .1688  & .1084  & .2041  & .1504  & .2599  \\
    +RDiffBR & .0720  & .0554  & .0928  & .0716  & .1027  & .0794  & .1060  & .0812  & .1088  & .0834  & .0681  & .1203  & .0926  & .1531  & .1026  & .1788  & .1241  & .2243  & .1611  & .2669  \\
    \textbf{\%improv.} & \textbf{15.36} & \textbf{12.87} & \textbf{5.75} & \textbf{4.78} & \textbf{2.83} & \textbf{2.82} & \textbf{1.31} & \textbf{1.64} & \textbf{0.60} & \textbf{0.32} & \textbf{14.80} & \textbf{11.40} & \textbf{8.01} & \textbf{13.15} & \textbf{5.95} & \textbf{9.37} & \textbf{9.89} & \textbf{14.48} & \textbf{2.69} & \textbf{7.12} \\[-1pt]
    \midrule
    MultiCBR & .0672  & .0588  & .1080  & .0905  & .1297  & .1064  & .1343  & .1101  & .1381  & .1133  & .0989  & .1782  & .1147  & .1841  & .1305  & .2092  & .1476  & .2376  & .1625  & .2592  \\
    +RDiffBR & .0694  & .0608  & .1094  & .0916  & .1313  & .1080  & .1350  & .1111  & .1385  & .1139  & .1066  & .1889  & .1217  & .1945  & .1375  & .2164  & .1524  & .2436  & .1633  & .2598  \\
    \textbf{\%improv.} & \textbf{3.29} & \textbf{3.31} & \textbf{1.23} & \textbf{1.16} & \textbf{1.20} & \textbf{1.47} & \textbf{0.53} & \textbf{0.97} & \textbf{0.27} & \textbf{0.58} & \textbf{6.04} & \textbf{7.83} & \textbf{5.63} & \textbf{6.10} & \textbf{3.44} & \textbf{5.40} & \textbf{2.51} & \textbf{3.19} & \textbf{0.25} & \textbf{0.50} \\[-1pt]
    \midrule
    BundleGT & .0444  & .0352  & .0690  & .0517  & .0954  & .0724  & .0899  & .0653  & .0960  & .0707  & .0732  & .1281  & .0923  & .1556  & .1093  & .1761  & .1217  & .2134  & .1579  & .2585  \\
    +RDiffBR & .0456  & .0361  & .0701  & .0530  & .0970  & .0740  & .0909  & .0660  & .0970  & .0712  & .0863  & .1512  & .1082  & .1770  & .1315  & .2077  & .1414  & .2346  & .1707  & .2733  \\
    \textbf{\%improv.} & \textbf{2.68} & \textbf{2.58} & \textbf{1.64} & \textbf{2.34} & \textbf{1.61} & \textbf{2.10} & \textbf{1.02} & \textbf{1.04} & \textbf{1.10} & \textbf{0.71} & \textbf{18.03} & \textbf{17.88} & \textbf{13.75} & \textbf{17.20} & \textbf{17.97} & \textbf{20.35} & \textbf{9.91} & \textbf{16.21} & \textbf{5.73} & \textbf{8.05} \\ [-1pt]
\bottomrule
%\addlinespace[0.15cm] 

\end{tabular}

\caption{Recall@20 and NDCG@20 performance comparison between backbone models and our RDiffBR on four datasets ($\rho$=-4, -2, 0, 2, 4), while $\rho <0$ indicates $Z_\mathbf{test} = Z_\mathbf{train\&val}\setminus \Delta^-$, $\rho>0$ indicates adding set $Z_\mathbf{test} = Z_\mathbf{train\&val}\cup \Delta^+$, and $\rho =0$ indicates $Z_\mathbf{train\&val}$ invariability. Improv. indicates the relative improvement of a backbone model+RDiffBR over the itself. The performance of K$ = \{10, 40, 80\} $ is consistent with  K$ = 20 $. }
  \label{tab:perform}%
\end{table*}%

%%%%%%%%%%%%%%%%%%%%%%%%%%%%%%%%%%%%%%%%%%%%%%%%%%%%%%%%%%%%%%%%%%%%%
$\bullet$ \textbf{BR Backbones.}
We implement RDiffBR on six representative BR models with official codes released by their authors.
\textbf{BGCN}~\cite{BGCN} utilizes graph convolutional networks to separately learn the representations of bundle view and item view.
\textbf{HyperMBR}~\cite{HyperMBR} models two views in hyperbolic space to learn their accurate representations.
\textbf{MIDGN}~\cite{MIDGN} refines the user-item/bundle-item graph of different intents from multiple views.
\textbf{CrossCBR}~\cite{CrossCBR} improves two separate views by constrastive learning.
\textbf{MultiCBR}~\cite{MultiCBR} first fuses the multi-view representations into a unified one, and then adopts a contrastive loss on the unified representations.
\textbf{BundleGT}~\cite{BundleGT} learns the strategy-aware user and bundle representation in bundle recommendation. %There are still some earlier research outcomes (for instance, BundleNet~\cite{BundleNet}) that have been studied to be less effective than the aforementioned methods. Therefore, we do not take them into consideration.

\begin{figure}[!t]
    \centering
    \includegraphics[width=1\linewidth]{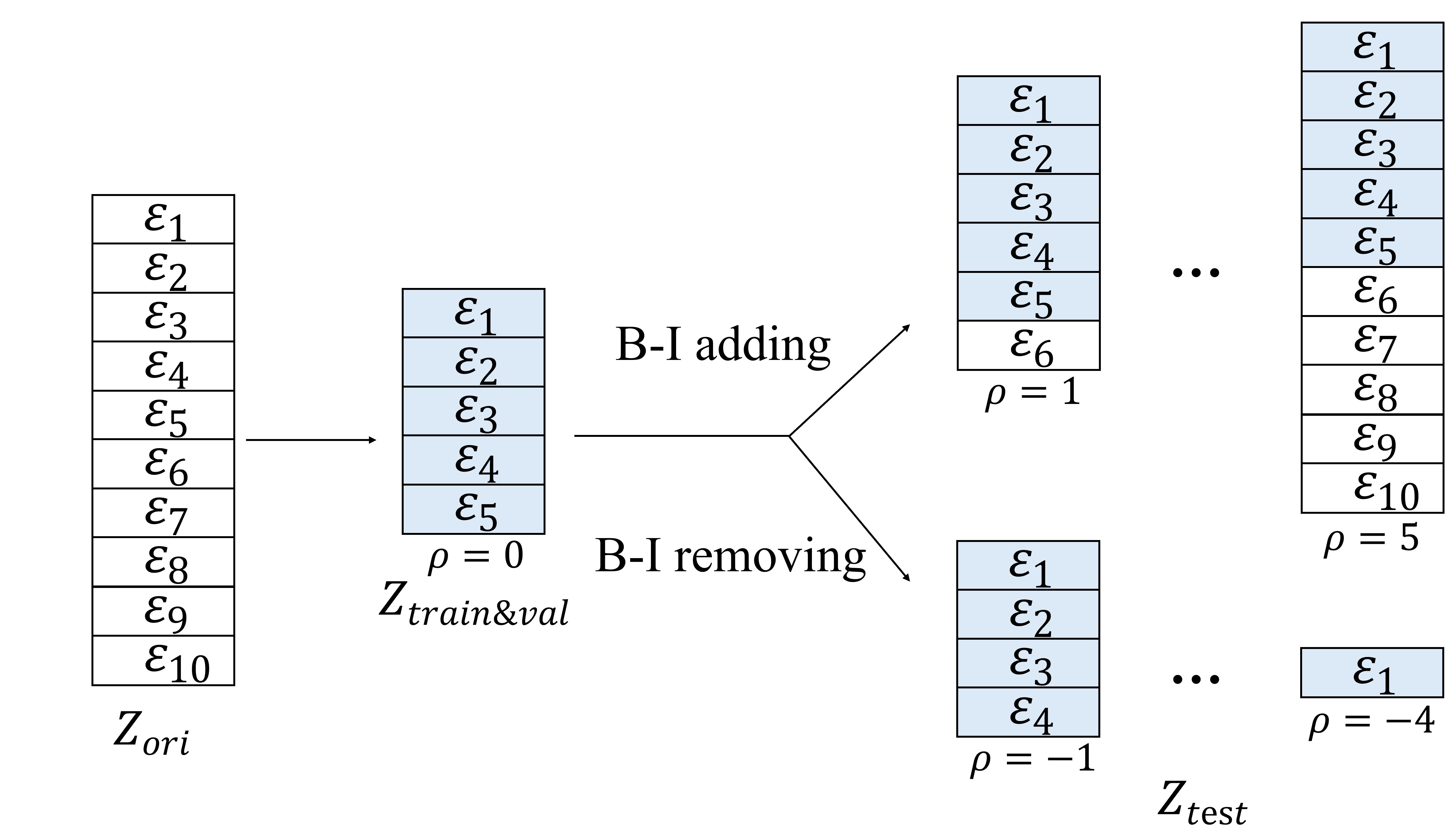}
    \caption{The processing of the B-I affiliation to match the scenario of B-I dynamic variability.}
    \label{fig processing dataset}
\end{figure}
$\bullet$ \textbf{Experimental Preprocess and Metrics.}
According to previous works~\cite{CrossCBR}~\cite{mealrec+}, the training, validation and testing sets of user-bundle interactions $X$ are randomly splited with specific ratio. It is not feasible to simply add new bundles in bundle-item affiliations $Z$ under the B-I variability assumptions. Therefore, we adjust existing bundle datasets to match this scenario: Let $\{{\mathcal{E}_1, \mathcal{E}_2,…, \mathcal{E}_{10}}\}$ denote a uniform partition of $Z$ into ten disjoint subsets. Bundle models are trained and validated on $Z_\mathrm{train\&val}=\cup_{i=1}^{5}\mathcal{E}_{i}$. In testing phase, we design $Z_\mathrm{test}$ by adjusting $Z_\mathrm{train\&val}$  with adding or removing subsets:
\begin{equation}Z_\mathrm{test}=\begin{cases}Z_{\mathrm{train\&val}}\cup\left(\mathcal{\cup}_{i=6}^{5+\rho}\mathcal{E}_{i}\right)&\mathrm{if}~\rho>0\\Z_\mathrm{{train\&val}}&\mathrm{if ~\rho=0}\\Z_{\mathrm{train\&val}}\setminus\left(\mathcal{\cup}_{i=6-|\rho|)}^{5}\mathcal{E}_{i}\right)&\mathrm{if ~\rho< 0}\end{cases}\end{equation}
Where $\cup $ indicates set union and $\setminus$ indicates set difference. $\rho \in\{-4,-3,-2,-1,0,1,2,3,4,5\}$ reflects level of B-I variation. $\rho >0$ indicates adding set $\Delta^+$ to $Z_\mathrm{test}$, $\rho <0$ indicates removing set $\Delta^-$ from $Z_\mathrm{test}$  and $\rho =0$ indicates $Z_\mathrm{test}$ invariability. The well-trained bundle models predict user-bundle interactions $X$ on $Z_\mathrm{test}$. The processing of the B-I affiliation is shown in Figure~\ref{fig processing dataset}. In this way, we can control the degree of B-I affiliatons variation and simulate the scenario of B-I dynamic variability. Recall@K and NDCG@K are used as the evaluation metrics, where K $\in \{10,20, 40,80\}$.

$\bullet$ \textbf{Hyperparameter Settings}
 The hyperparameters of six backbone models are set by following their publications and RDiffBR is optimized with the Adam optimizer with a learning rate of 0.001. The time steps $T$ and sampling steps $T'$ are set to values ranging from [1,200]. The loss balance factor $\lambda$ is chosen from (0,5), while the residual connection weight $\delta$ is selected from (0,1). The number of MLP layers is selected within {1, 2, 4, 8}. Its hidden sizes are from [8,4096]. Details of other hyperparameters can be found in our released code. All the models are trained using two GTX 4090.

\begin{figure*}[!t]
  \centering
  \includegraphics[width=\linewidth]{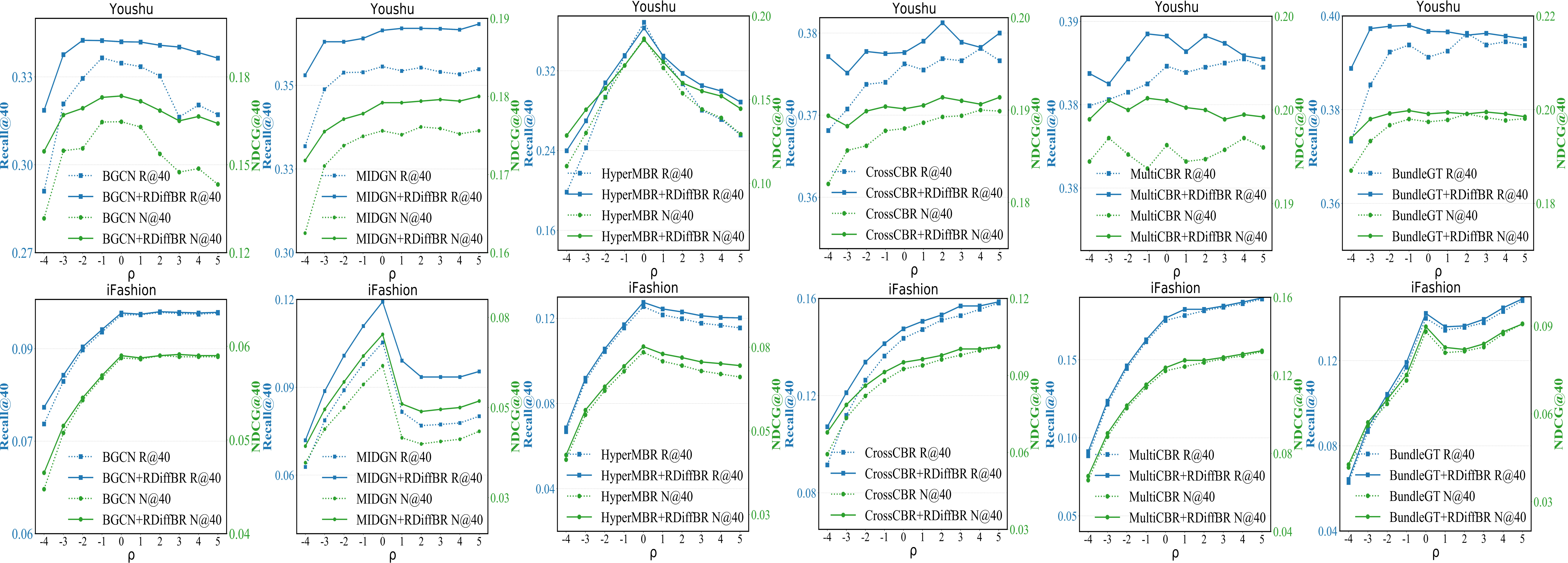}
    % \vspace{-4mm}
  \caption{Recall@40 and NDCG@40 performance comparisons of the backbone models and our RDiffBR under different levels of B-I variations in terms of Recall@40 and NDCG@40. The performance of K$ = \{10, 20, 80\} $ is consistent with  K$ = 40 $. }
   % \vspace{-2mm}
  \label{figPerformance}
\end{figure*}

\subsection{Overall Performance (RQ1)}

% We present the results of Recall@20 and NDCG@20 for five levels of B-I variations in Table~\ref{tab:perform}. We show the results of Recall@40 and NDCG@40 on two datasets for ten levels of B-I variations in Figure~\ref{figPerformance}. The following observations can be made: %%More results on XXX are in Appendix.X.%%
The following observations are from Table~\ref{tab:perform}. %and Figure~\ref{figPerformance}:%
\\$\bullet$ \textbf{The performance of BR backbones when meeting item-level dynamic variability.} When $\rho<0$ ($\mathbf{Z_{test}}$ removes $\Delta^-$), the performance of bundle models is generally lower than that under $\rho=0$ in most cases. The greater the reduction in B-I, the more significant the performance degradation. When $\rho>0$ ($\mathbf{Z_{test}}$ adds $\Delta^+$), some models exhibit performance decline, others maintain comparable performance, while a few show improvement. This indicates that BR models possess capability to handle B-I variations to some extent, yet they tend to fail under drastic reductions of B-I. Notably, HyperMBR performs the worst, where both B-I adding and removing frequently lead to performance deterioration. This phenomenon may stem from hyperbolic space embeddings being more sensitive to distance errors compared to Euclidean space. In contrast, models like MIDGN, CrossCBR, and MultiCBR exhibit relatively stable performance, potentially attributed to their adoption of contrastive learning that leverages multi-view information through cooperative learning. Additionally, BundleGT demonstrates enhanced adaptability for dynamic B-I variations by  modeling B-I affiliation via attention mechanisms.
\\$\bullet$ \textbf{The performance of RDiffBR dealing with item-level dynamic variability.}
Our experiments reveal that RDiffBR consistently enhances the performance of six backbone BR models across four datasets from different domains under different B-I variation levels. When $\rho < 0$ ($\mathbf{Z_{test}}$ removes $\Delta^-$), The performance of BR models+RDiffBR outperform BR models with the improvement  up to 22.15\% at most. When $\rho > 0$ ($\mathbf{Z_{test}}$ adds $\Delta^+$), RDiffBR also enhances the performance of BR models, reaching up to 24.83\% at most. In some cases, RDiffBR can even enable BR models to outperform its performance under $\rho = 0$, which highlights the powerful generative capability of RDiffBR. Furthermore, performance improvements is observed even under $\rho = 0$, which demonstrates RDiffBR has certain effect even in the B-I remain scenario. Overall, RDiffBR can enhance the performance of BR models with different architectures and metric spaces under different variation scenarios. This remarkable performance improvement can be attributed to RDiffBR’s generation capability, which reconstructs item-level bundle embeddings in B-I change scenarios to better align with bundle themes rather than overfitting to the B-I affiliations observed during training.

\subsection{Ablation Study (RQ2)}

To assess whether residual connections contribute to improving model performance, we set a variant \textbf{RDiffBR w/o R that omits the residual connection} in Eq.\eqref{7} and conduct a comprehensive ablation study on three BR models across three datasets, with the results shown in Table~\ref{tab:Ablation}. 
\begin{table}[h]
  \centering
  
% \vspace{-1mm}
  \Huge
        \resizebox{\linewidth}{!}{
      
    \begin{tabular}{lcccccc}
    \toprule
    \multicolumn{1}{c}{\multirow{2}[4]{*}{Method($\rho=-3$)}} & \multicolumn{2}{c}{Youshu } & \multicolumn{2}{c}{NetEase} & \multicolumn{2}{c}{MealRec$^+_H$} \\
\cmidrule{2-7}          & R@20   & N@20   & R@20   & N@20   & R@20   & N@20 \\
    \midrule
 BGCN  & $\underline{0.2214}$  & $\underline{0.1280 }$ & 0.0621  & $\underline{0.0328}$  & 0.1962  & 0.1122  \\
    BGCN+RDiffBR & \textbf{0.2366 } & \textbf{0.1398 } & \textbf{0.0647 } & \textbf{0.0336 } & \textbf{0.2159 } & $\underline{0.1240 }$ \\
    BGCN+RDiffBR w/o R& 0.2158  & 0.1151  & $\underline{0.0624}$  & 0.0323  & $\underline{0.2118}$  & \textbf{0.1245 } \\
    \midrule
    CrossCBR & $\underline{0.2742}$  & 0.1623  & 0.0767  & 0.0413  & 0.1229  & 0.0720  \\
    CrossCBR+RDiffBR & \textbf{0.2789 } & \textbf{0.1646 } & \textbf{0.0810 } & \textbf{0.0433 } & $\underline{0.1329}$  & $\underline{0.0798}$  \\
    CrossCBR+RDiffBR w/o R& 0.2740  & $\underline{0.1628}$  & $\underline{0.0782}$  & $\underline{0.0418}$  & \textbf{0.1508 } & \textbf{0.0825 } \\
    \midrule
    BundleGT &$ \underline{0.2836}$  & $\underline{0.1656}$  & $\underline{0.0838}$  & $\underline{0.0438}$  & 0.1398  & 0.0841  \\
    BundleGT+RDiffBR & \textbf{0.2908 } & \textbf{0.1693 } & \textbf{0.0864 } & \textbf{0.0450 } & $\underline{0.1616 }$ & $\underline{0.0982}$  \\
    BundleGT+RDiffBR w/o R& 0.2519  & 0.1454  & 0.0520  & 0.0267  & \textbf{0.2313 } & \textbf{0.1416 } \\

    \bottomrule
    \end{tabular}% 
    }

  \centering
          \resizebox{\linewidth}{!}{
    \begin{tabular}{lcccccc}
    \toprule
    \multicolumn{1}{c}{\multirow{2}[4]{*}{Method($\rho=3$)}} & \multicolumn{2}{c}{Youshu } & \multicolumn{2}{c}{NetEase} & \multicolumn{2}{c}{MealRec$^+_H$} \\
\cmidrule{2-7}          & R@20   & N@20   & R@20   & N@20   & R@20   & N@20 \\
    \midrule
      BGCN  & 0.2129  & 0.1197  & $\underline{0.0632}$  & $\underline{0.0336 }$ & $\underline{0.2298}$  & $\underline{0.1350}$  \\
    BGCN+RDiffBR & \textbf{0.2394 } & \textbf{0.1374 } & \textbf{0.0660 } & \textbf{0.0343 } & \textbf{0.2390 } & \textbf{0.1404 } \\
    BGCN+RDiffBR w/o R& $\underline{0.2337} $ &$\underline{ 0.1229}$  & 0.0627  & 0.0321  & 0.2089  & 0.1221  \\
    \midrule
    CrossCBR & $\underline{0.2795 }$ &$\underline{ 0.1653} $ & $\underline{0.0801} $ &$ \underline{0.0435  }$& $\underline{0.2395} $ & $\underline{0.1295}$  \\
    CrossCBR+RDiffBR & \textbf{0.2816 } & \textbf{0.1664 } & \textbf{0.0827 } & \textbf{0.0445 } & \textbf{0.2474 } & \textbf{0.1440 } \\
    CrossCBR+RDiffBR w/o R& 0.2763  & 0.1639  & 0.0798  & 0.0426  & 0.1843  & 0.1052  \\
    \midrule
    BundleGT & $\underline{0.2858 }$ & $\underline{0.1691}$  &$\underline{ 0.0871}$  & $\underline{0.0466}$  & $\underline{0.2394}$ & 0.1416  \\
    BundleGT+RDiffBR & \textbf{0.2925 } & \textbf{0.1711 } & \textbf{0.0895 } & \textbf{0.0477 } & \textbf{0.2589 } & \textbf{0.1593 } \\
    BundleGT+RDiffBR w/o R& 0.2519  & 0.1454  & 0.0520  & 0.0267  & 0.2312  &$\underline{ 0.1417 }$ \\
    \bottomrule
    \end{tabular}%
    }
    \caption{Ablation study where bold indicates the best method, and underlined indicates the second best method($\rho=-3,3$).}

  \label{tab:Ablation}%
       % \vspace{-5mm}
\end{table}%

\begin{figure}[h]
    \centering
    \includegraphics[width=1\linewidth]{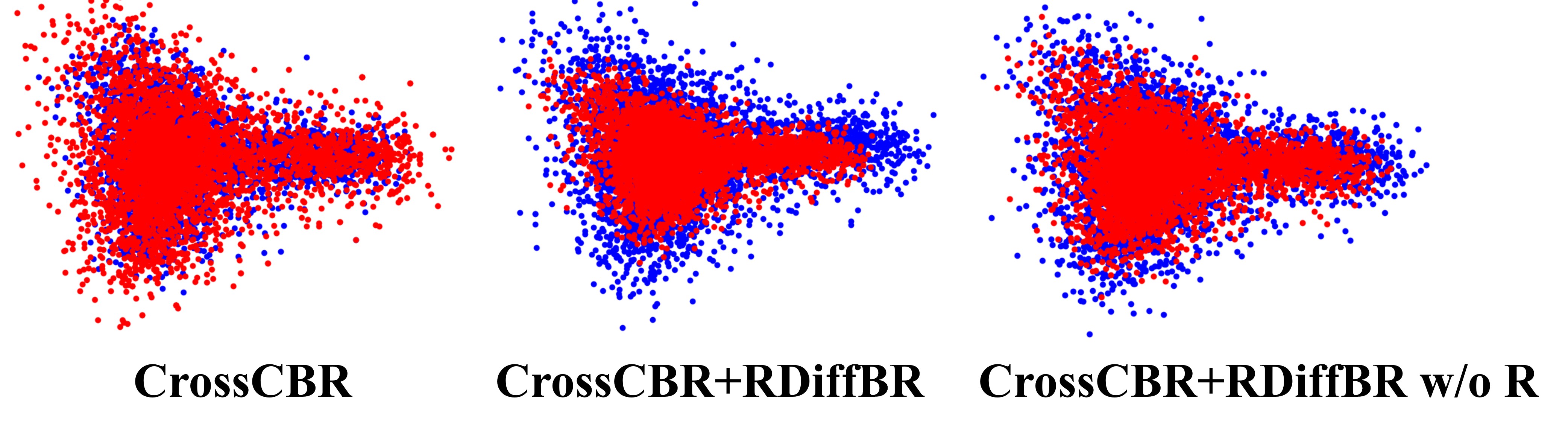}
    \vspace{-5mm}
    \caption{Blue is distribution $E^{IL}_B$ under $\rho=0$. Red is distribution $E^{IL}_B$  under $\rho=-3$.}
    \label{figVisualization1}
 \vspace{2mm}
    \centering
    \includegraphics[width=1\linewidth]{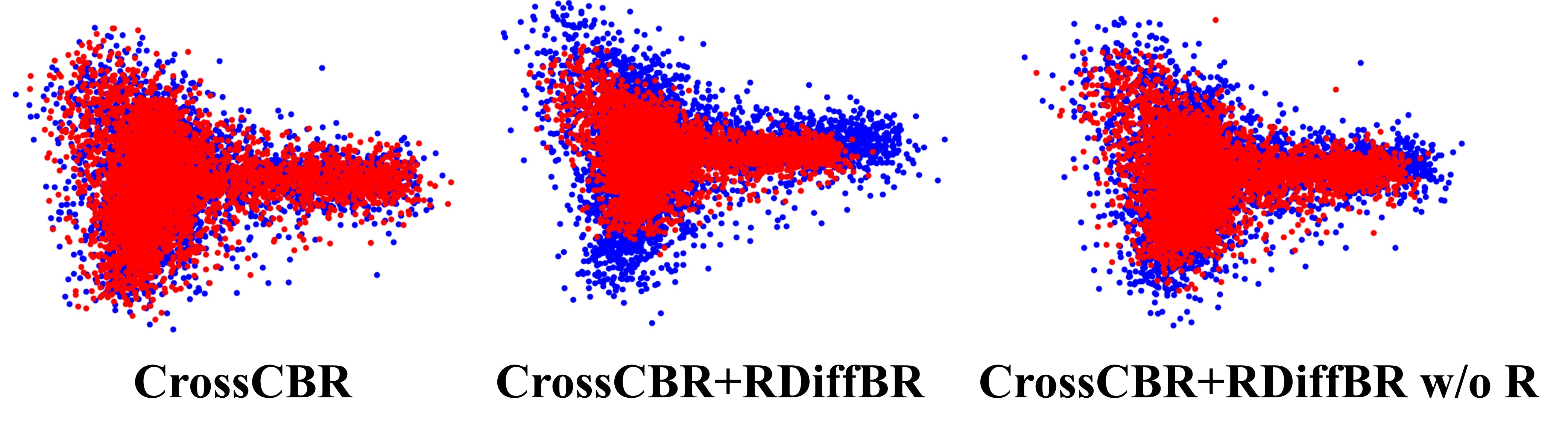}
     % \vspace{-1mm}
    \caption{Blue is distribution $E^{IL}_B$ under $\rho=0$. Red is distribution $E^{IL}_B$  under $\rho=3$. }
      % \vspace{-2.5mm}

    \label{figVisualization2}

\end{figure}

From this table, we can draw the following conclusions: In most cases, the performance of using RDiffBR is superior to that of not using it or using RDiffBR w/o R. Using RDiffBR w/o R may lead to worse performance than RDiffBR, and in some cases it may even be worse than only using backbone BR model. Figure~\ref{figVisualization2} shows the visualization results of CrossCBR, CrossCBR+RDiffBR and CrossCBR+RDiffBR w/o R under B-I varation levels $\rho=3$ for item-level bundle embeddings $E^{IL}_B$ in Youshu dataset. It can be observed that $E^{IL}_B$ generated by CrossCBR after B-I change is more dispersed than original $E^{IL}_B$ and more compact $E^{IL}_B$ distribution  which retains the basic shape can be obtained after using CrossCBR+RDiffBR. This indicates that $E^{IL}_B$ of the BR model cannot effectively represent the bundle theme distribution due to the B-I change and RDiffBR enables the model to obtain more bundle information at the item-level rather than overfitting to the existing b-I affiliations. While using CrossCBR+RDiffBR w/o R, $E^{IL}_B$ distribution changes little so the effect is limited, which is consistent with the experimental results obtained in Table~\ref{tab:Ablation}. Overall, RDiffBR outperforms backbone models, highlighting the effectiveness of RDiffBR. Residual connection plays an important role in RDiffBR, and merely using a diffusion model cannot generate better item-level bundle embeddings (The performance of $\rho=-3$ is consistent with $\rho=3$, see details in extended version).

Although different B-I affiliations are used during model testing, the bundle theme remains, which means items within a bundle remain to follow the similar distribution before and after the change. This can also be verified by Figure~\ref{figVisualization2} that the shapes of two $E^{IL}_B$ distributions under different B-I variation $\rho$ are similar. Therefore, with using RDiffBR, our task can be approached without considering the Out-Of-Distribution (OOD) where the distribution of the training data and the test data is inconsistent. 
In extreme cases where the distribution of $E^{IL}_B$ changes completely, the problem becomes equivalent to cold-start bundle recommendation~\cite{cold_start_1,cold_start_2} that the original bundle id cannot be directly associated.

\subsection{Impact of Hyperparameters}

\begin{figure}[!t]
    \centering
    \includegraphics[width=1\linewidth]{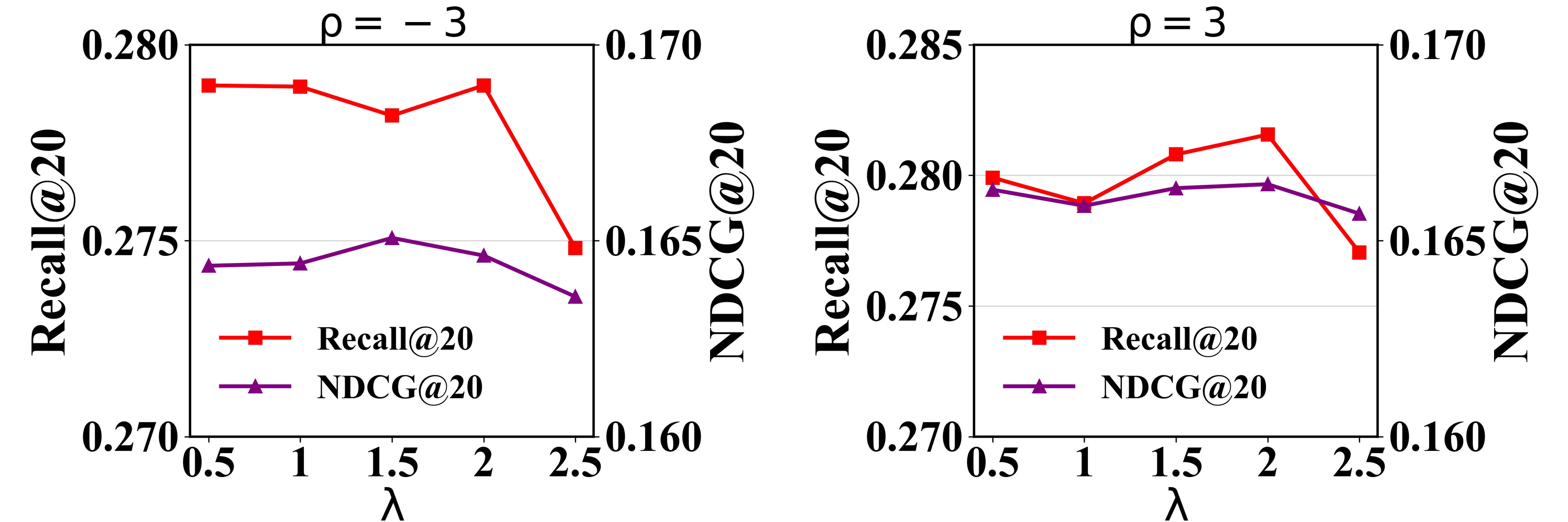}
    % \vspace{-7mm}
    \caption{The impact of loss balance factor $\lambda$ %in CrossCBR with RDiffBR for Youshu in two B-I variation levels
    .}
    \label{figloss balance factor}
    \centering
    \includegraphics[width=1\linewidth]{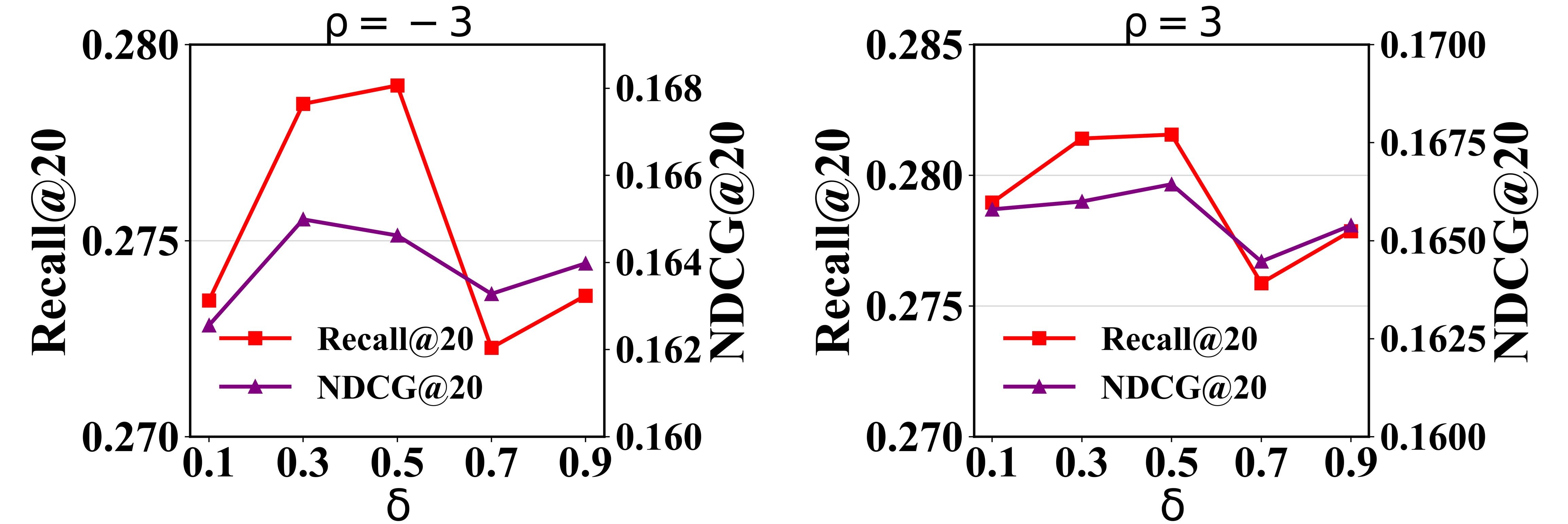}
     % \vspace{-7mm}
    \caption{The impact of residual weight $\delta$ % in CrossCBR with RDiffBR for Youshu in two B-I variation levels
    .}
    \label{figresidual weight}

    % \centering
    % \includegraphics[width=1\linewidth]{12.jpg} 
    % \vspace{-7mm}
    % \caption{The impact of dimensionality of time step embedding $d$ %in CrossCBR with RDiffBR for Youshu in two B-I variation levels
    % .}
    % \label{figdimensionality of time step}

\end{figure}

In order to achieve a more detailed understanding, two key hyperparameters are studied, i.e., the loss balance factor $\lambda$ and the residual weight $\delta$ . The results of our hyperparameter study are presented in Figure~\ref{figloss balance factor}-\ref{figresidual weight}.% by adjusting them within a certain range.
\\$\bullet$\textbf{ Loss balance factor $\lambda$.} The loss balance factor $r$ affects the performance of RDiffBR, which adjusts the emphasis between the recommendation task and the reconstruction task. A high value of $r$ may neglect the core recommendation and thereby compromise the overall performance.
\\$\bullet$\textbf{ Residual weight $\delta$.} The appropriate selection of residual weights is crucial for embedding generation, this is because too high or too low values will lead to a performance decline.
% \\$\bullet$\textbf{ Dimensionality of time step embedding $d$.} As the value of d increases, the model performance initially improves. This is because the time step embedding can flexibly adjust the denoising strategy according to different noise stages. However, when the value of d is excessively large, the input embedding will gradually be weakened during the process, thereby negatively affecting the model performance.

\subsection{Case Study}
\begin{figure}[!t]
    \centering
    \includegraphics[width=1\linewidth]{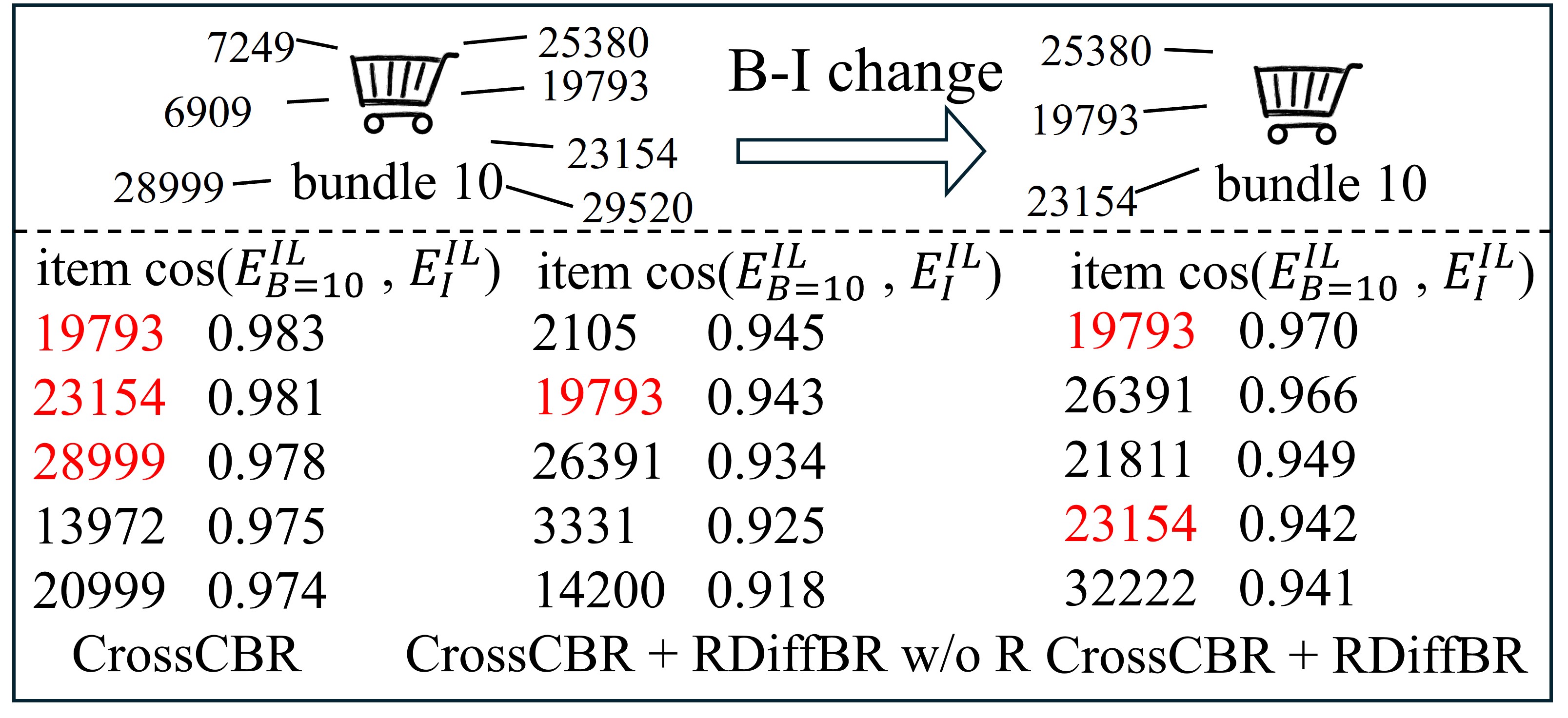}
    \caption{A case study for Youshu under $\rho=-3$. The highlighted items are those before the bundle change. }
    \label{fig:case study1}
% \vspace{-0.5cm}
\end{figure}
\begin{figure}[!t]
    \centering
    \includegraphics[width=1\linewidth]{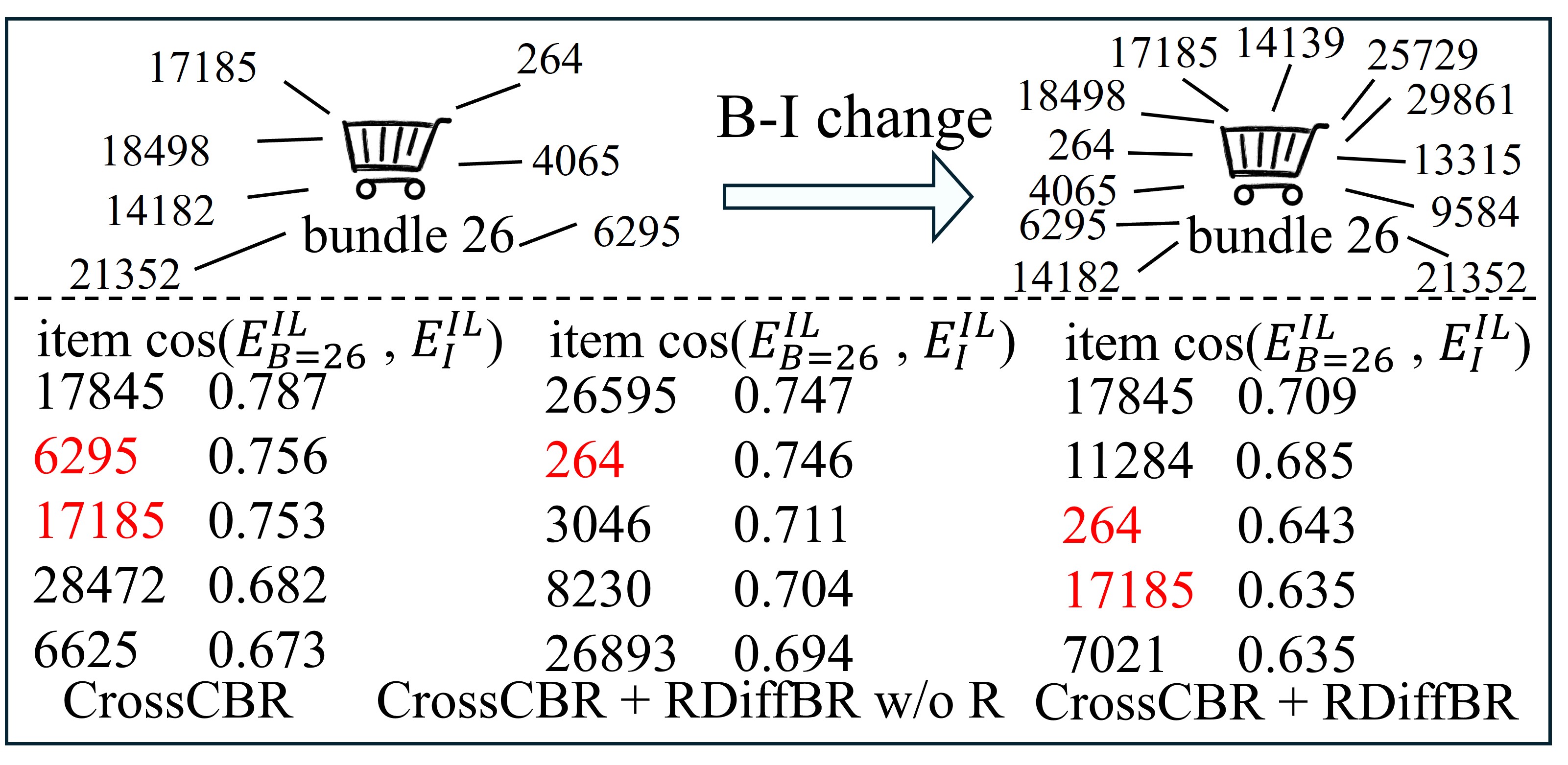}
    \caption{A case study for Youshu under $\rho=3$. The highlighted items are those before the bundle change.}
    \label{fig:case study2}
\end{figure}
In this section, we analyze the real data instances as show in Figure~\ref{fig:case study1}. We randomly select a bundle from Youshu and calculate cosine similarity between item-level bundle embeddings $E^{IL}_B$ and all items under B-I varation levels $\rho=-3$. We listed the top 5 items with the highest similarity scores and their corresponding similarity values. It can be seen that the similarity of items (marked in red) that appeared within the bundle will be higher when only using CrossCBR, resulting in insufficient generalization ability. When using RDiffBR w/o R, the model cannot well capture the main items of the bundle. When using RDiffBR, the model can not only express the main items of the bundle but also introduce other items based on the theme to enhance the generalization ability.

\subsection{Training Efficiency Study (RQ3)}
In order to evaluate the additional computing time of RDiffBR, we compare the one-epoch training time (seconds) of using BR model and BR model+RDiffBR on four datasets (The training rounds are the same for both modes). We record five consecutive training epochs and averaged them to obtain the one-epoch training time, as shown in Table~\ref{tab:training time}. It shows that  RDiffBR only increased training time about 4\%.
% \begin{table}[!h]
    
%   \centering
%   \samll
%   % \vspace{0.25cm}

%     \begin{tabular}{l|c|c|c|c|c}
%     \toprule
%           Dataset&      \multicolumn{1}{c}{Youshu}   &\multicolumn{1}{c}{NetEase}  &\multicolumn{1}{c}{iFashion}    & \multicolumn{1}{c}{MealRec$^+_H$} \\

%     \midrule
%     BGCN  &        1.46  &        23.5 &68.3 &1.07 \\
%     BGCN+RDiffBR       &1.51(+3.42\%)     & 24.5(+4.25\%)   & 71.4(+4.53\%)  & 1.09(+1.86\%)   \\
%     \midrule
%     CrossCBR &        1.06  &        8.70 &45.4 & 1.10   \\
%     CrossCBR+RDiffBR &        1.11(+4.71\%)         & 9.10(+4.59\%)    & 46.2(+1.76\%)  & 1.14(+3.63\%)   \\
%     \midrule
%     BundleGT &             22.7    &       426.2&  118.4& 1.23\\
%     BundleGT+RDiffBR &        23.3(+2.64\%)          &       434.8(+2.01\%)   & 121.1(+2.28\%)  &  1.29(+4.87\%) \\
%     \bottomrule
%     \end{tabular}%
 
%     \caption{The statistics of one-epoch training time(seconds) for three backbone models and using RDiffBR on them.}
%   \label{tab:training time}%
% \end{table}%
%%%%%%%%%%%%%%%%%%%%%%%%%%%%%%%%%%%%%%%%%
\begin{table}[t]

\centering
% \small

\label{tab:statistics}
\begin{tabularx}{0.47\textwidth}{lcccc}
\toprule
Dataset          &Youshu  &NetEase  &iFashion &MealRec$^+_H$   \\ 
\hline

    BGCN          & 1.46  &        23.5 &68.3 &1.07 \\
    +RDiffBR       &1.51     & 24.5   & 71.4  & 1.09  \\
    Extra cost       &3.42\%     & 4.25\%  & 4.53\%  & 1.86\%   \\
    \hline
    CrossCBR &        1.06  &        8.70 &45.4 & 1.10   \\
    +RDiffBR &        1.11        & 9.10   & 46.2 & 1.14  \\
    Extra cost   & 4.71\%         & 4.59\%    & 1.76\%  & 3.63\%\\
    \hline
    BundleGT &             22.7    &       426.2&  118.4& 1.23\\
    +RDiffBR &        23.3         &       434.8  & 121.1  &  1.29 \\
    Extra cost &       2.64\%          &      2.01\%   & 2.28\%  &  4.87\%\\
   \bottomrule

\end{tabularx}
\caption{The statistics of one-epoch training time (seconds) for three backbone models and using RDiffBR on them.}
\label{tab:training time}%
\end{table}
%%%%%%%%%%%%%%%%%%%%%%%%%%%%%%%%%%%%%%%%%%%

From the perspective of time complexity, taking lightweight backbone CrossCBR as an example, the time complexity of graph learning is $\mathcal{O}(|E|Kd)$, where $|E|$ is the number of all edges in U-B and U-I graphs, $K$ is the number of propagation layers, $d$ is the embedding size. However, the diffusion process in RDiffBR costs $\mathcal{O} (|B|d) + \mathcal{O} (|B|d)d’n$, where $|B|$ is the number of bundles, $d’$ is the embedding size of MLP in approximator, $n$ is MLP layer number. It shows that $|E|$ is much greater than $|B|$ (e.g., in iFashion $|E|=1679708 >> |B|=27694$). Therefore, our RDiffBR only increases training time about $4\%$ on two GTX 4090.

\section{Conclusion and Future Work}
In this work, to address the issue of item-level dynamic variability, we propose RDiffBR, a model-agnostic generative framework that can help BR models adapt to this scenario. %Given the item-level recommendation model of the bundle, RDiffBR actively injects Gaussian noise into these embeddings and iteratively removes the noise during the reverse process. To guide the reverse process, we designed a residual approximator module. In the inference stage, we used this module to generate new item-level embeddings. 
A number of experiments show RDiffBR enables backbone BR models to achieve significant performance improvement under dynamic changes of B-I. Future research will consider extending RDiffBR as a plug-and-play module for well-performing pre-trained bundle recommenders and evaluating its applicability to cold-start bundle recommendation.

\section{Acknowledgments}
This work is partially supported by NSFC, China (No.62276196), the research grant (RGPIN-2020-07157) from the Natural Science and Engineering Research Council (NSERC) of Canada and York Research Chairs (YRC) program. The authors also gratefully appreciate the anonymous reviewers for their valuable comments and constructive suggestions that greatly helped to improve the quality of the paper.

\bibliography{aaai2026}

\end{document}